\documentclass[aps,prl,showpacs,amsmath,amssymb,twocolumn,10pt]{revtex4-1} 
\usepackage{bbm}
\usepackage[dvips]{graphicx}

\newcommand\RR{{\mathbbm{R}}}
\newcommand\CC{{\mathbbm{C}}}

\newcommand\ii{{\mathrm{i}}}

\DeclareMathOperator{\Tr}{Tr}

\begin{document} 

\title{Nonequivalence of Ensembles for Long-Range Quantum Spin Systems in Optical Lattices} 

\author{Michael Kastner} 
\email{kastner@sun.ac.za} 
\affiliation{National Institute for Theoretical Physics (NITheP), Stellenbosch 7600, South Africa} 
\affiliation{Institute of Theoretical Physics,  University of Stellenbosch, Stellenbosch 7600, South Africa}

\date{\today}
 
\begin{abstract}
Motivated by the anisotropic long-range nature of the interactions between cold dipolar atoms or molecules in an optical lattice, we study the anisotropic quantum Heisenberg model with Curie-Weiss-type long-range interactions. Absence of a heat bath in optical lattice experiments suggests a study of this model within the microcanonical ensemble. The microcanonical entropy is calculated analytically, and nonequivalence of microcanonical and canonical ensembles is found for a range of anisotropy parameters. From the shape of the entropy it follows that the Curie-Weiss Heisenberg model is indistinguishable from the Curie-Weiss Ising model in canonical thermodynamics, although their microcanonical thermodynamics differs. Qualitatively, the observed features of nonequivalent ensembles are expected to be relevant for long-range quantum spin systems realized in optical lattice experiments.
\end{abstract}


\maketitle 

Cold dipolar gases have been in the very focus of experimental and theoretical research recently \cite{Gries_etal05,Ni_etal08}. In particular, dipolar gases in optical traps have been suggested as laboratory realizations of lattice spin models where the coupling parameters can be tuned freely, allowing for the realization of many Hamiltonians of interest in condensed matter physics \cite{Micheli_etal06}.

After switching off the cooling in such an experiment, total energy and number of atoms are conserved to a very good degree. As a consequence, a statistical description of such a lattice spin model should make use of the microcanonical ensemble. For systems with short-range interactions, the choice of the statistical ensemble is typically of minor importance and could be considered a finite-size effect: differences between, say, microcanonical and canonical expectation values are known to vanish in the thermodynamic limit of large system size, and the various statistical ensembles become equivalent \cite{Ruelle}. In the presence of long-range interactions this is in general not the case, and microcanonical and canonical approaches can lead to different thermodynamic properties even in the infinite-system limit \cite{TouElTur04}. In the astrophysical context, nonequivalence of ensembles and the importance of microcanonical calculations have long been known for gravitational systems \cite{LynWood68,Thirring}.

In condensed matter physics, most systems are coupled to an environment, and therefore the canonical or grandcanonical ensembles are the ones that appropriately describe the experimental situation of interest. Moreover, screening effects lead in general to interactions which are effectively of short range, and hence equivalence of ensembles usually can be taken for granted. As a consequence, calculations of thermodynamic quantities can be done in the ensemble that is the most convenient one, which appears to be the canonical or grandcanonical, but never the microcanonical, one. Owing to these facts, little is known about such systems in the microcanonical ensemble. Only recently, a number of toy models, consisting of long-range coupled classical spin variables, has been studied (see \cite{CamDauxRuf09} for a review). The study of these strongly simplified but analytically solvable models has been very fruitful towards the aim of understanding general dynamical and thermodynamical properties of classical systems with long-range interactions.

Much less is known about the peculiarities of quantum spin systems with long-range interactions, and, in particular, about equivalence or nonequivalence of ensembles in this context. It is the aim of this Letter to contribute towards the understanding of such systems, with a focus onto the microcanonical setting as encountered in experiments with dipolar gases in optical lattices.

To this purpose, we study the anisotropic quantum Heisenberg model with Curie-Weiss-type long-range interactions in the microcanonical ensemble. Such Curie-Weiss-type interactions, where each spin is interacting with every other at equal strength, are clearly an idealization of the actual interactions of dipolar atoms which decay like $r^{-3}$ with the interparticle distance $r$. However, it is known that Curie-Weiss-type models faithfully reproduce many properties of algebraically decaying long-range interactions qualitatively, and to some extent even quantitatively \cite{BisChayCraw06,Chayes09}.

In this Letter the result of an exact, analytic calculation of the thermodynamic limit of the microcanonical entropy of the anisotropic quantum Heisenberg model with Curie-Weiss-type interactions is reported. Depending on the choice of the anisotropy parameters in the Hamiltonian operator, a concave entropy function is found in some cases, and a nonconcave one in others. Correspondingly, equivalence of the microcanonical and the canonical ensemble holds in the first case, but not in the second. 

The relevance of the reported results is twofold. First, the observation of nonequivalent ensembles in long-range quantum spin systems demonstrates that, under the experimental conditions realized in cold dipolar gases in optical traps, the choice of the statistical ensemble is of paramount importance. Consequently, a statistical interpretation of the results of such experiments has to go beyond usual canonical thermodynamics. In particular, differences between microcanonical and canonical expectation values do not diminish in importance with increasing system size. This is in sharp contrast to microcanonical computations for ideal Bose gases in traps \cite{GrossHolt96}, where equivalence of ensembles holds in the thermodynamic limit. Second, the reported calculation also illustrates that cold dipolar gases in optical traps are excellent laboratory systems in which long-range effects like the nonequivalence of statistical ensembles or the negativity of microcanonical response functions can possibly be tested. Since these effects occur only for a certain range of values of the anisotropy parameters, it is of particular importance that coupling constants (and therefore anisotropy parameters) in cold atom experiments can be tuned with a high level of control, rendering such systems an ideal laboratory for the study of these fundamental issues of thermostatistical physics.

%------------------------------------------------
{\em Anisotropic quantum Heisenberg model.---} The model consists of $N$ spin-$1/2$ degrees of freedom, each of which is interacting with every other at equal strength. The corresponding Hilbert space $\mathcal{H}=(\CC^2)^{\otimes N}$ is the tensor product of $N$ copies of the spin-$1/2$ Hilbert space $\CC^2$, and the Hamilton operator is given by
\begin{equation}\label{eq:Hamiltonian}
H_h=-\frac{1}{2N}\sum_{k,l=1}^N \!\left(\lambda_1 \sigma_k^1 \sigma_l^1 + \lambda_2 \sigma_k^2 \sigma_l^2 + \lambda_3 \sigma_k^3 \sigma_l^3\right) - h\sum_{k=1}^N \sigma_k^3.
\end{equation}
The $\sigma_k^\alpha$ are operators on $\mathcal{H}$ and act like the $\alpha$ component of the Pauli spin-$1/2$ operator on the $k$th factor of the tensor product space $\mathcal{H}$, and like identity operators on all the other factors. The resulting commutation relation is
\begin{equation}
\bigl[\sigma_k^\alpha,\sigma_l^\beta\bigr]=2\ii\, \delta_{k,l}\,\epsilon_{\alpha\beta\gamma}\sigma_k^\gamma,\qquad\alpha,\beta\in\{1,2,3\},
\end{equation}
where $\delta$ denotes Kronecker's symbol and $\epsilon$ is the Levi-Civita symbol. $h$ is the strength of an external magnetic field orientated along the $3$ axis, and the constants $\lambda_1$, $\lambda_2$, and $\lambda_3$ determine the coupling strengths in the various spatial directions and allow us to adjust the degree of anisotropy. Note that it is explicitly shown in \cite{Micheli_etal06} that anisotropic quantum Heisenberg models are among the systems that can be engineered with cold polar molecules in optical lattices.

Special choices for the coupling constants in \eqref{eq:Hamiltonian} yield, for example, (a) the isotropic Heisenberg model, $\lambda_1=\lambda_2=\lambda_3$, (b) the Ising model, $\lambda_1=0=\lambda_2$, (c) the isotropic Lipkin-Meshkov-Glick model, $\lambda_1=\lambda_2$ and $\lambda_3=0$. For these special cases, the Hamiltonian \eqref{eq:Hamiltonian} can be expressed in terms of ${\mathbf S}^2$ and $S_3$, i.e., the square and the 3 component of a collective spin operator ${\mathbf S}=\sum{\mathbf \sigma}_k/2$. As a consequence, an angular momentum eigenbasis simultaneously diagonalizes $H_h$ and $S_3$ and the model can be solved by elementary means.

Here we consider the coupling constants $\lambda_1$, $\lambda_2$, and $\lambda_3$ to be nonnegative, but otherwise arbitrary, and in this case the model is known to display a transition from a ferromagnetic to a paramagnetic phase in the canonical ensemble. The exact expression for the canonical Gibbs free energy $g$ as a function of the inverse temperature $\beta=1/T$ \footnote{Boltzmann's constant is set to unity.} and the magnetic field $h$ is known for this model (and, in fact, for a larger class of systems) and can be found, for example, in \cite{PearceThompson75}.

%------------------------------------------------
{\em Microcanonical entropy.---} In thermodynamics, the energy $e$ is the variable conjugate to the inverse temperature $\beta$, and the magnetization $m$ is conjugate to $-\beta h$. So in the same way that $g(\beta,h)$ represents the fundamental quantity of the quantum Heisenberg model in the canonical ensemble, the microcanonical entropy $s(e,m)$ serves as a starting point for a microcanonical description in the thermodynamic limit. However, for a pair of variables $(e,m)$ corresponding to the pair of noncommuting operators $(H_0,M=2S_3)$, it is not even well established how to define a quantum microcanonical entropy, symbolically given by
\begin{equation}\label{eq:semdef}
s_N(e,m)=\frac{1}{N}\ln\Tr\left[\delta(Ne-H_0)\delta(Nm-M)\right].
\end{equation}
Note that the symbolic expressions make little mathematical sense and require some physically reasonable regularization. Extending a suggestion by Truong \cite{Truong74} to interacting systems, the definition
\begin{equation}\label{eq:sNemdef}
s_N(e,m)\!=\!\frac{1}{N}\ln\sum_{\bar{e},\bar{m}}\!\Tr\!\left[P_{H_0}(\bar{e}) P_M(\bar{m}) \right] \delta_\Delta(\bar{e}-e)\delta_\Delta(\bar{m}-m)
\end{equation}
seems to be physically reasonable, but difficult to apply in practice. Here, $\bar{e}$ and $\bar{m}$ denote eigenvalues of the operators $H_0/N$ and $M/N$, respectively. $\delta_\Delta$ is the characteristic function of the interval $[-\Delta,0]$, i.e.,  $\delta_\Delta(x)=1$ if $x\in[-\Delta,0]$, and zero otherwise. $P_{H_0}(\bar{e})$, $P_M(\bar{m})$ denote the eigenprojections of the operators $H_0$ and $M$ belonging to the eigenvalues $\bar{e}$ and $\bar{m}$, respectively.

We here report results for $s(e,m)=\lim_{N\to\infty}s_N(e,m)$ obtained by using a different regularization. The analytic calculation of $s$ uses, among others, some ingredients from a related canonical calculation by Tindemans and Capel \cite{TinCa74}. Details will be reported elsewhere, but the main steps of the calculation can be sketched as follows. (i) The deltas in \eqref{eq:semdef} are replaced by their Fourier integral representations. (ii) The Lie-Trotter formula is applied to separate the resulting exponential of the Hamiltonian into exponentials of the type $\exp\{c_\alpha (S_\alpha)^2\}$, $\alpha\in\{1,2,3\}$, with constants $c_\alpha$ and collective spin components $S_\alpha$. (iii) These exponentials are transformed into exponentials $\exp\{\tilde{c}_\alpha S_\alpha\}$ by applying the Hubbard-Stratonovich trick. The trade-off for steps (ii) and (iii) is a $3n+2$-dimensional integral, to be considered in the limit $n\to\infty$. The advantage, however, is that the Hilbert space trace of $\exp\{\tilde{c}_\alpha S_\alpha\}$ factorizes into traces over the single-spin Hilbert spaces $\CC^2$, which can be easily performed. (iv) The resulting high-dimensional complex integral can be solved in the thermodynamic limit $N\to\infty$, for example, by the method of steepest descent.

\begin{figure}\center
\vspace{1mm}
\parbox{3cm}{\includegraphics[width=29mm]{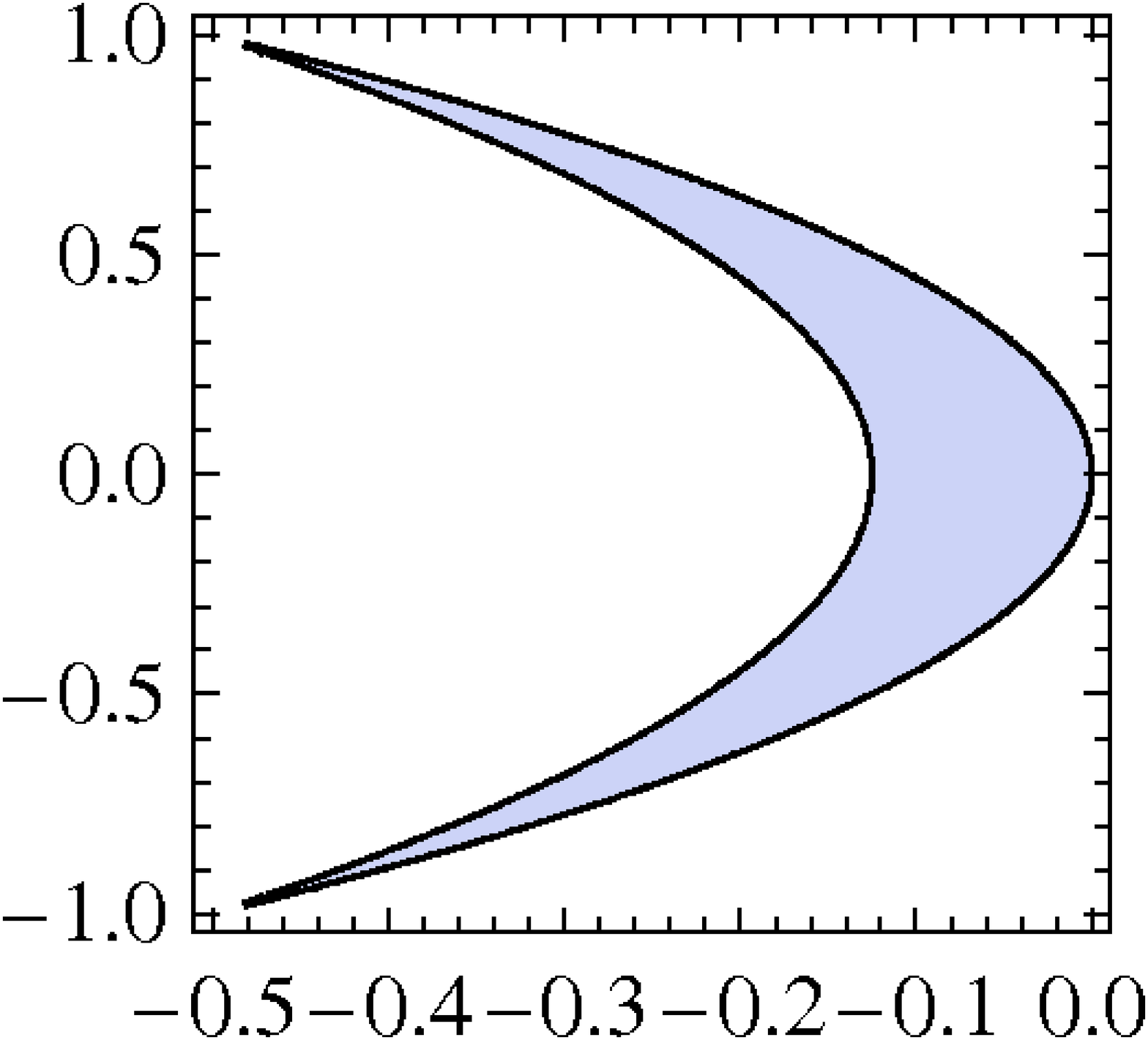}}
\hspace{6mm}
\parbox{4cm}{\includegraphics[width=37mm]{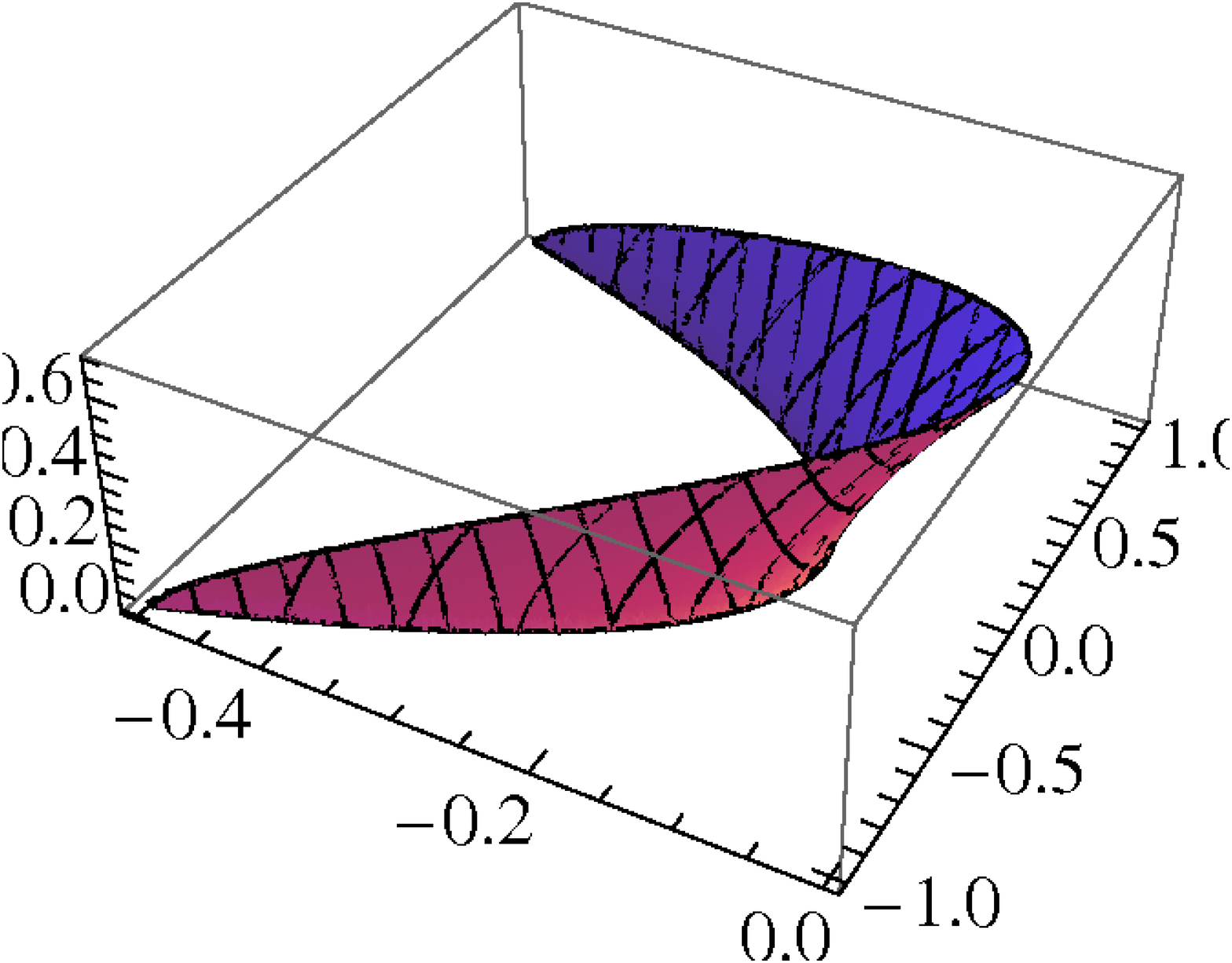}}
\newline
\parbox{3cm}{\includegraphics[width=29mm]{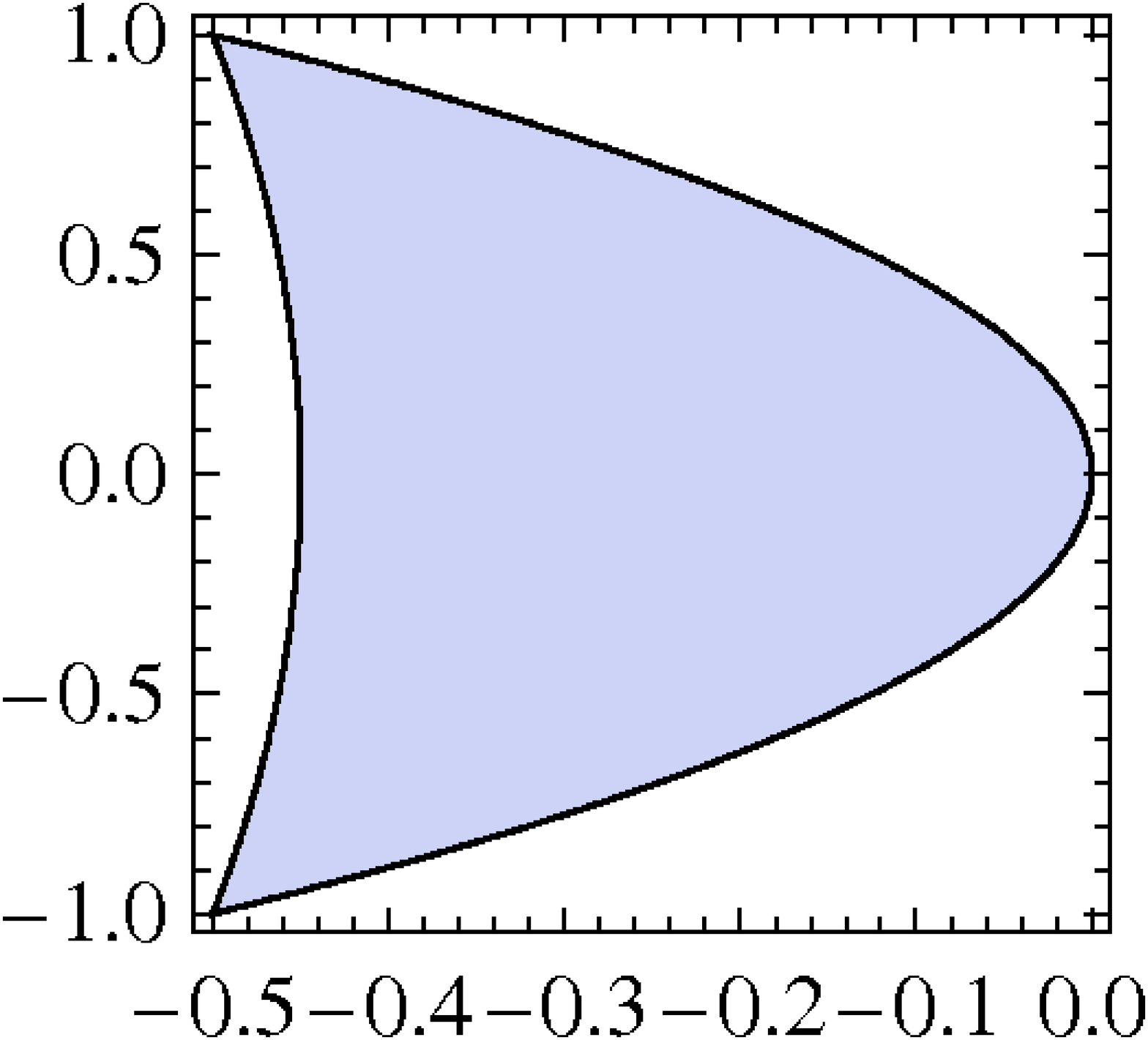}}
\hspace{6mm}
\parbox{4cm}{\includegraphics[width=37mm]{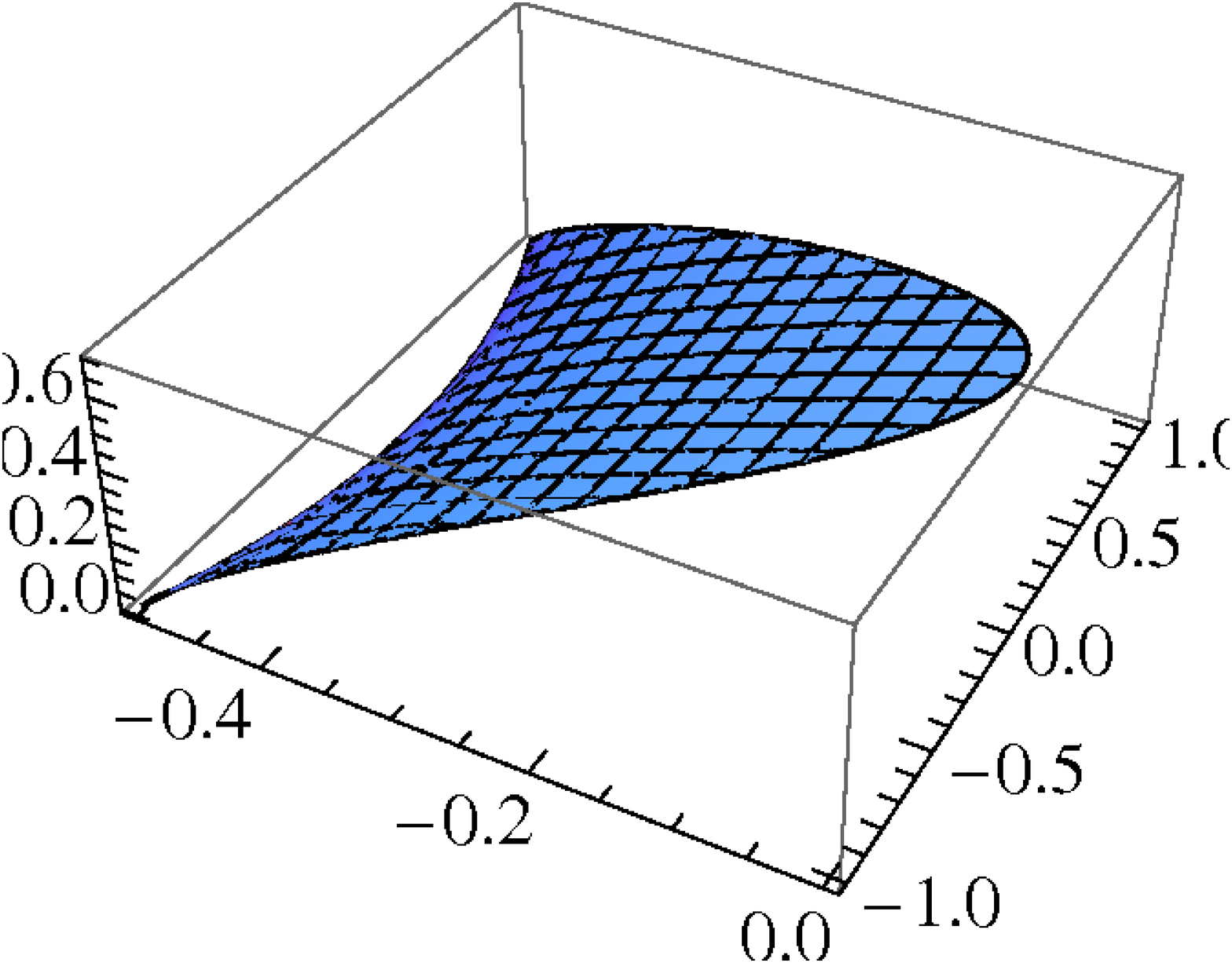}}
\newline
\parbox{3cm}{\includegraphics[width=29mm]{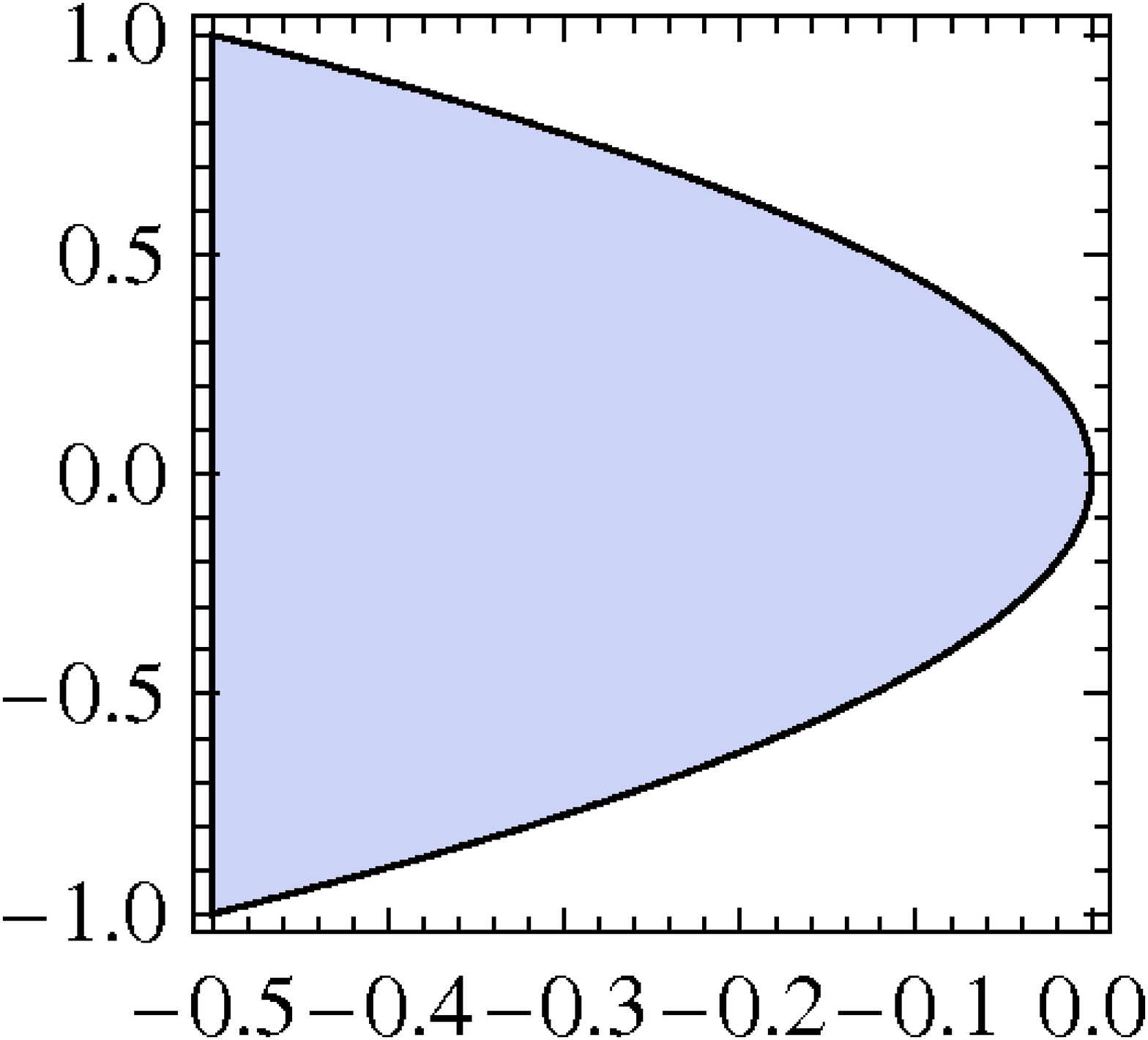}}
\hspace{6mm}
\parbox{4cm}{\includegraphics[width=37mm]{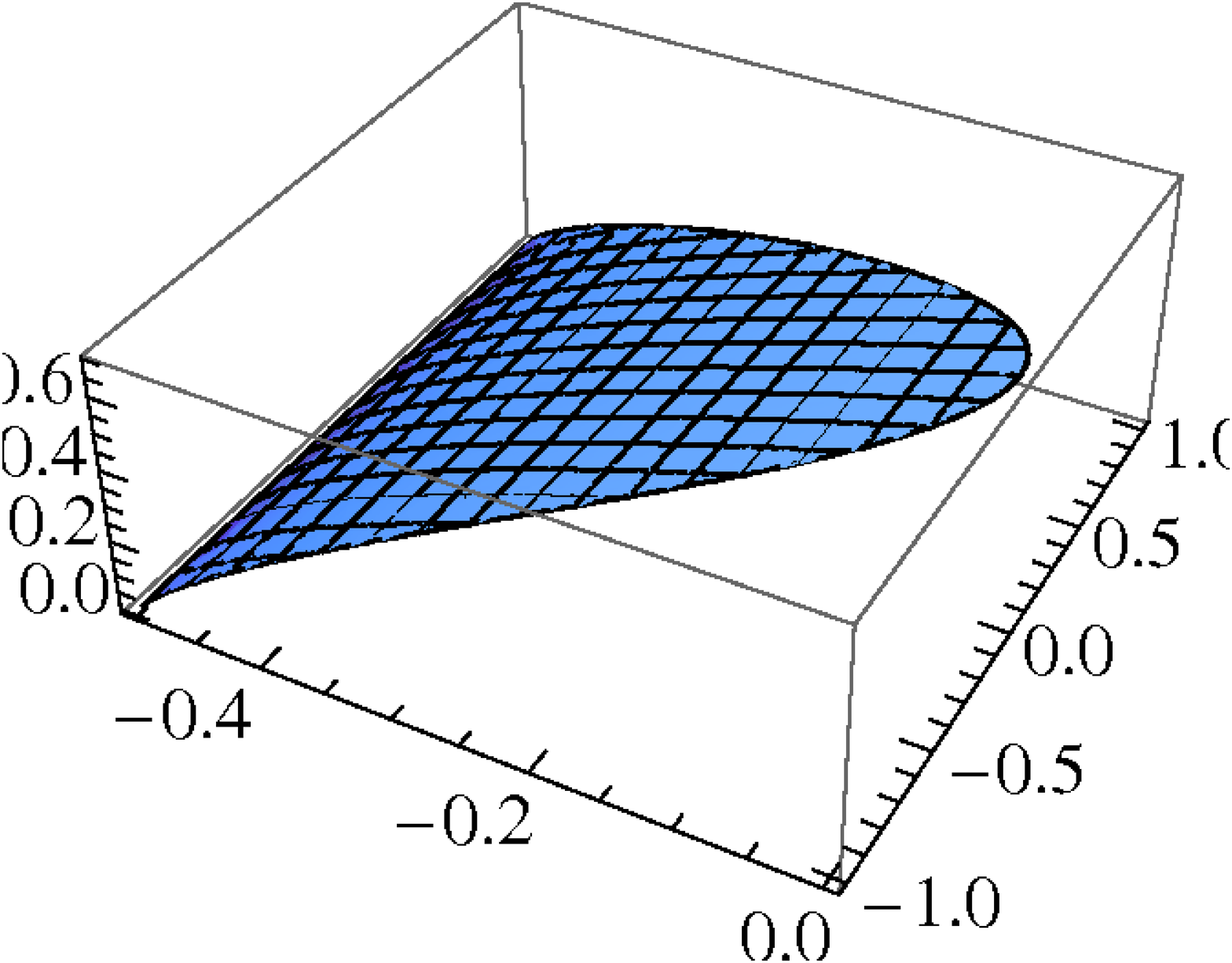}}
\newline
\parbox{3cm}{\includegraphics[width=29mm]{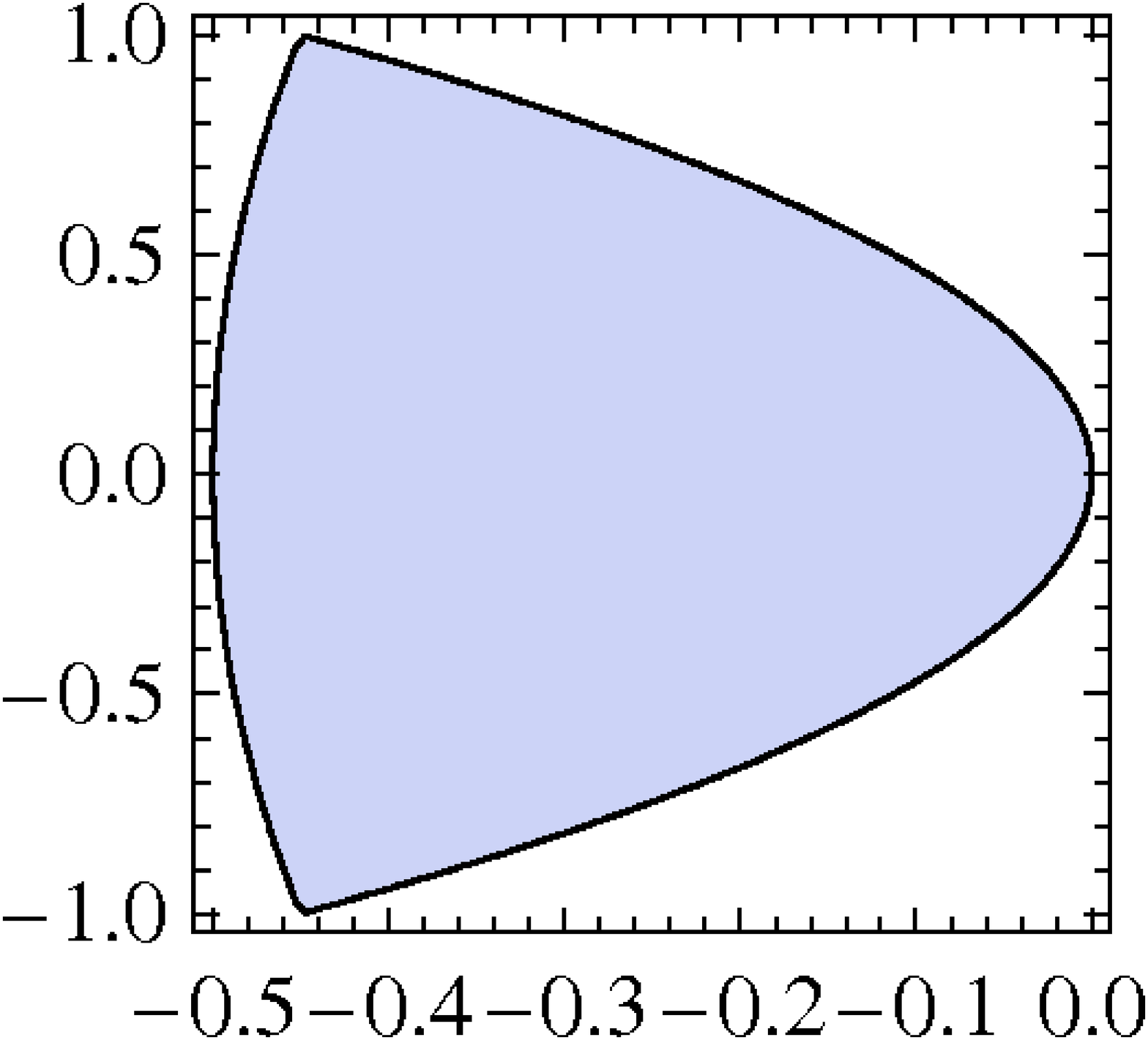}}
\hspace{6mm}
\parbox{4cm}{\includegraphics[width=37mm]{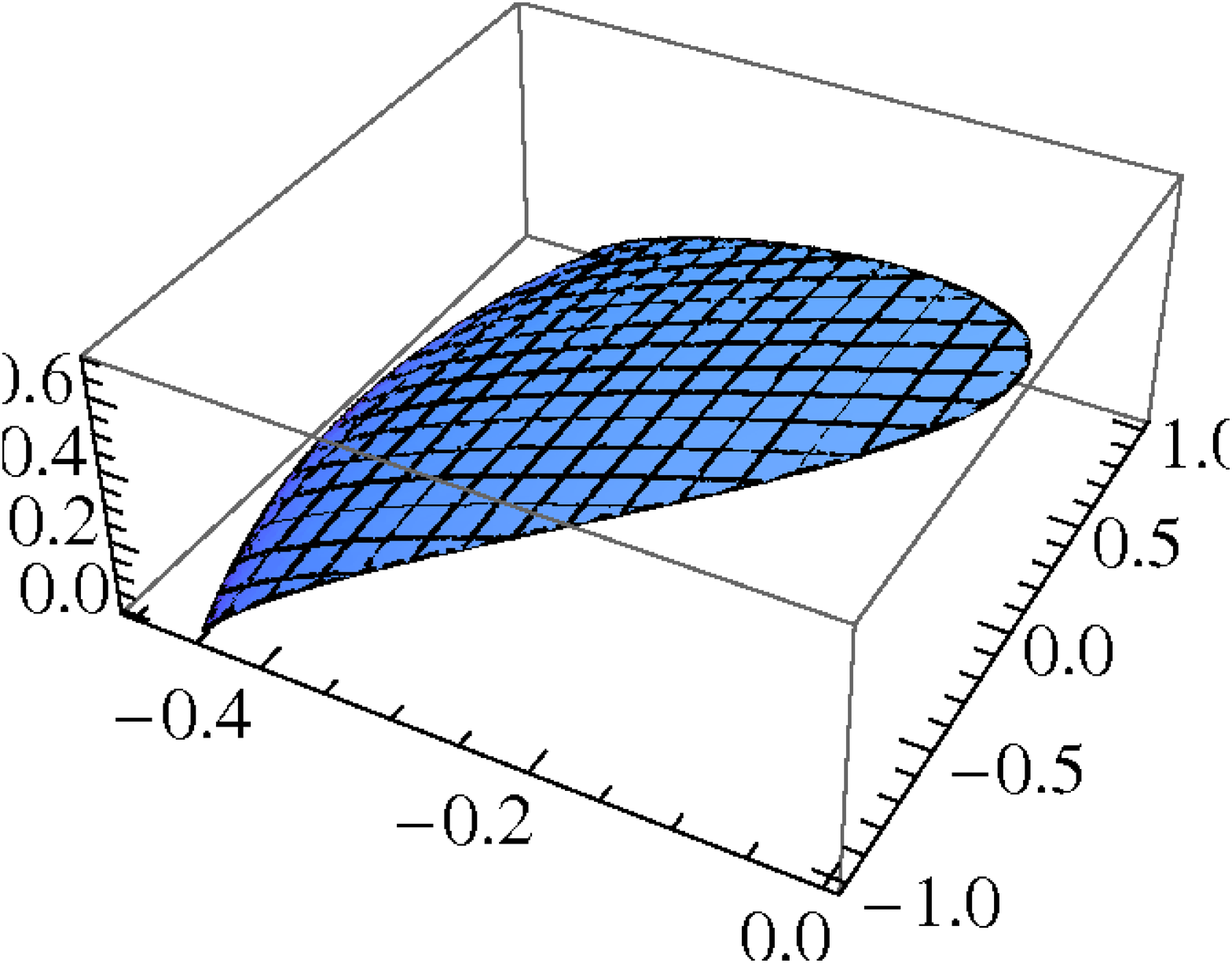}}
\newline
\parbox{3cm}{\includegraphics[width=29mm]{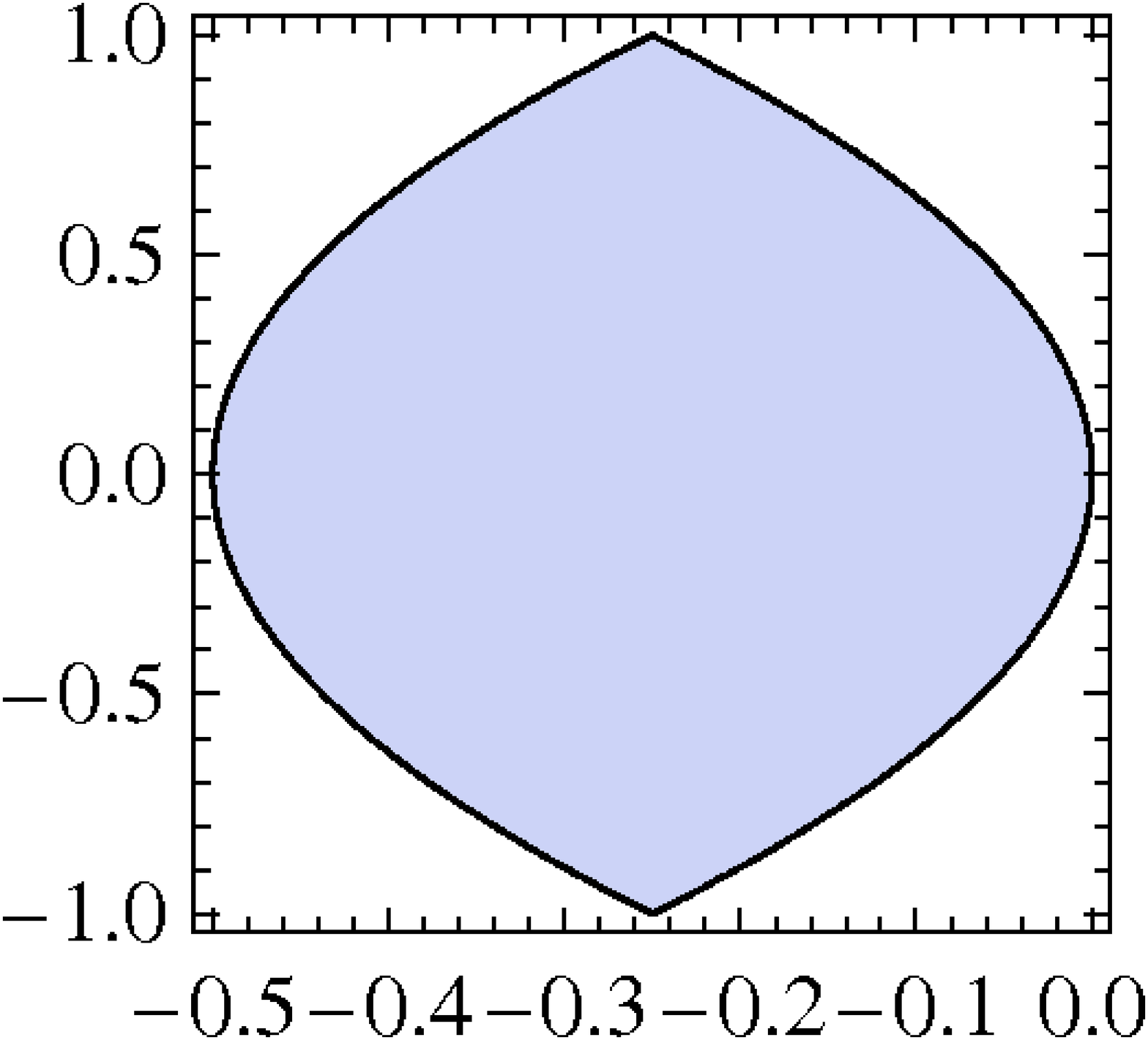}}
\hspace{6mm}
\parbox{4cm}{\includegraphics[width=37mm]{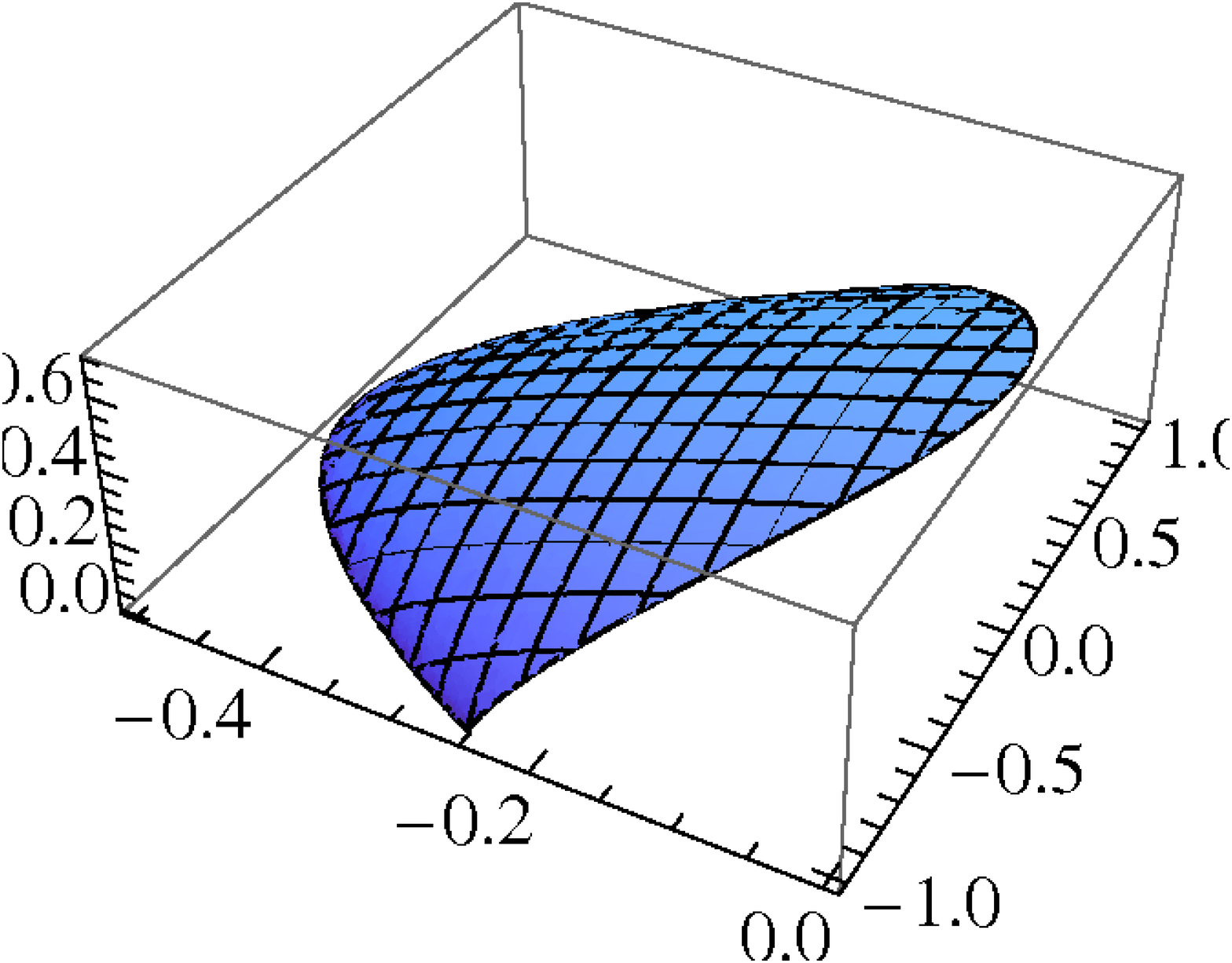}}
\newline
\parbox{3cm}{\includegraphics[width=29mm]{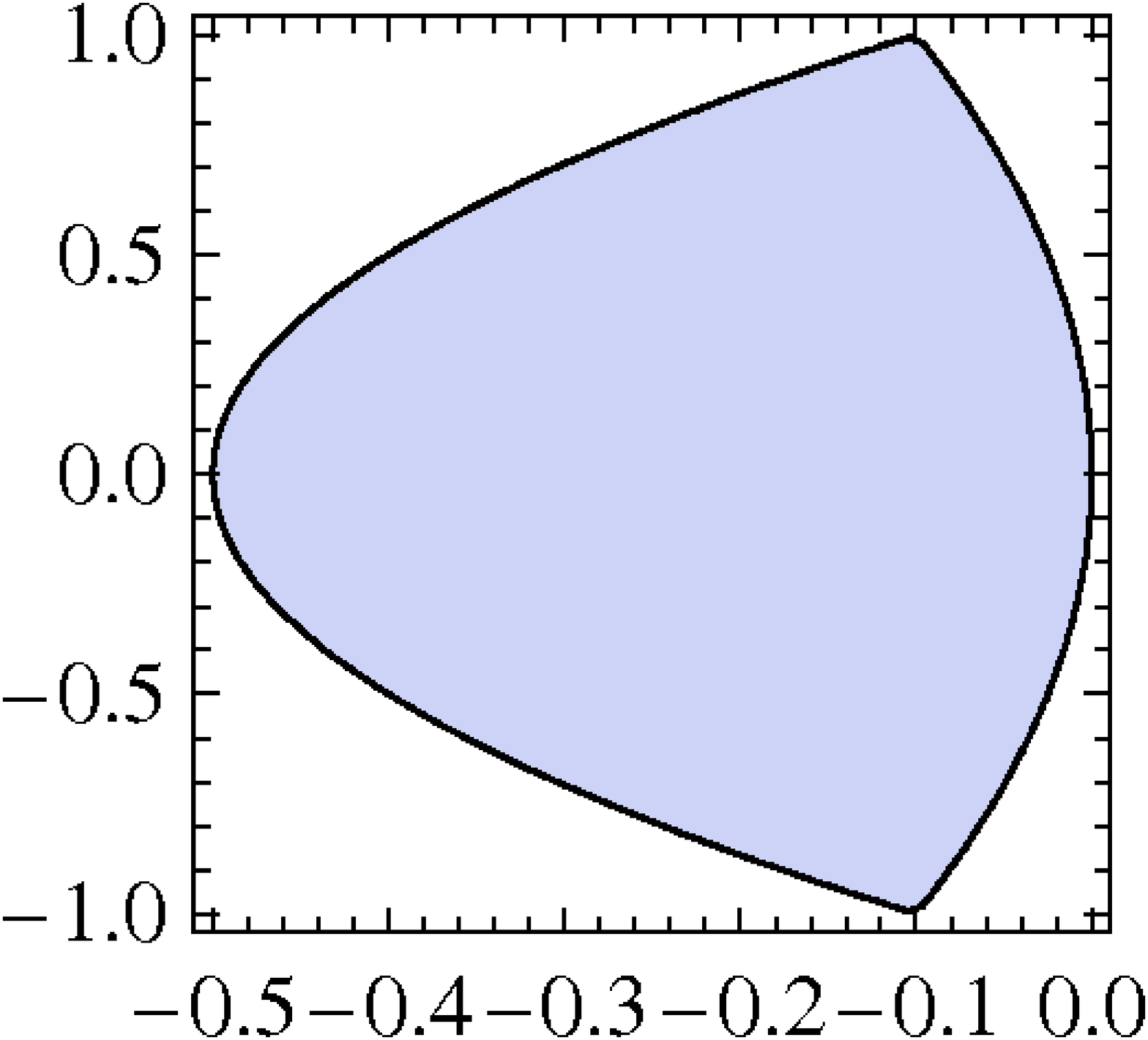}}
\hspace{6mm}
\parbox{4cm}{\includegraphics[width=37mm]{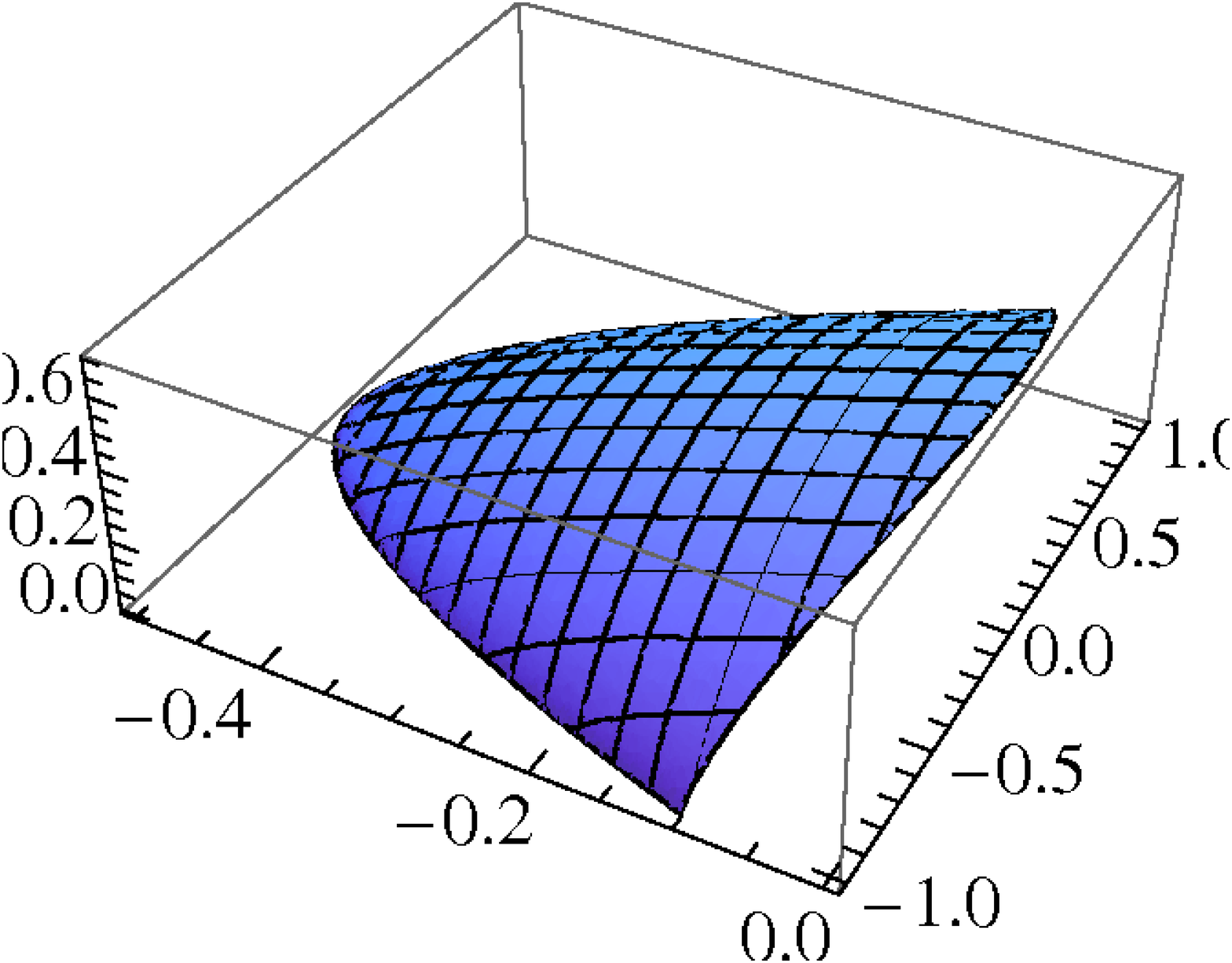}}
\newline
\parbox{3cm}{\includegraphics[width=29mm]{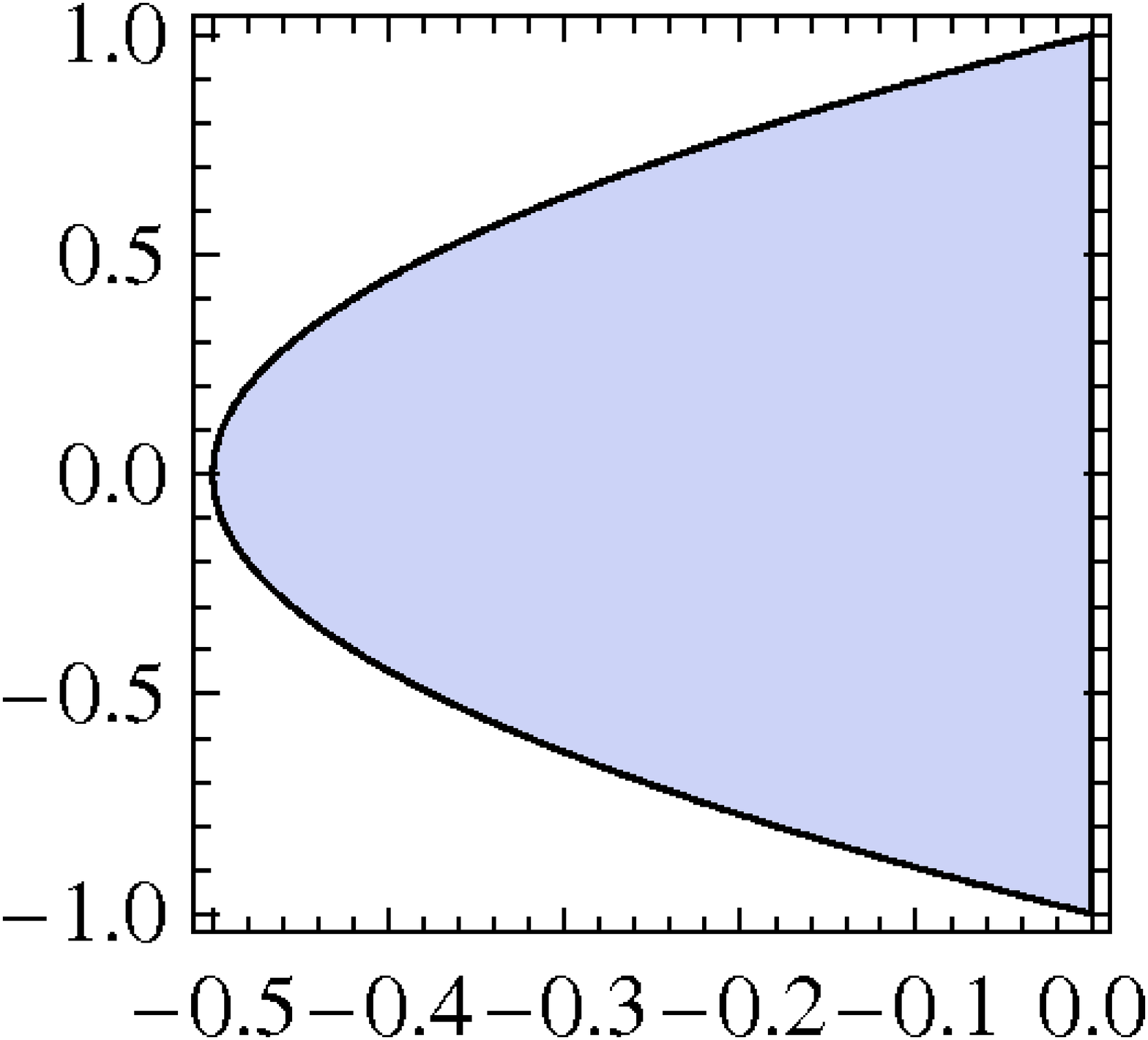}}
\hspace{6mm}
\parbox{4cm}{\includegraphics[width=37mm]{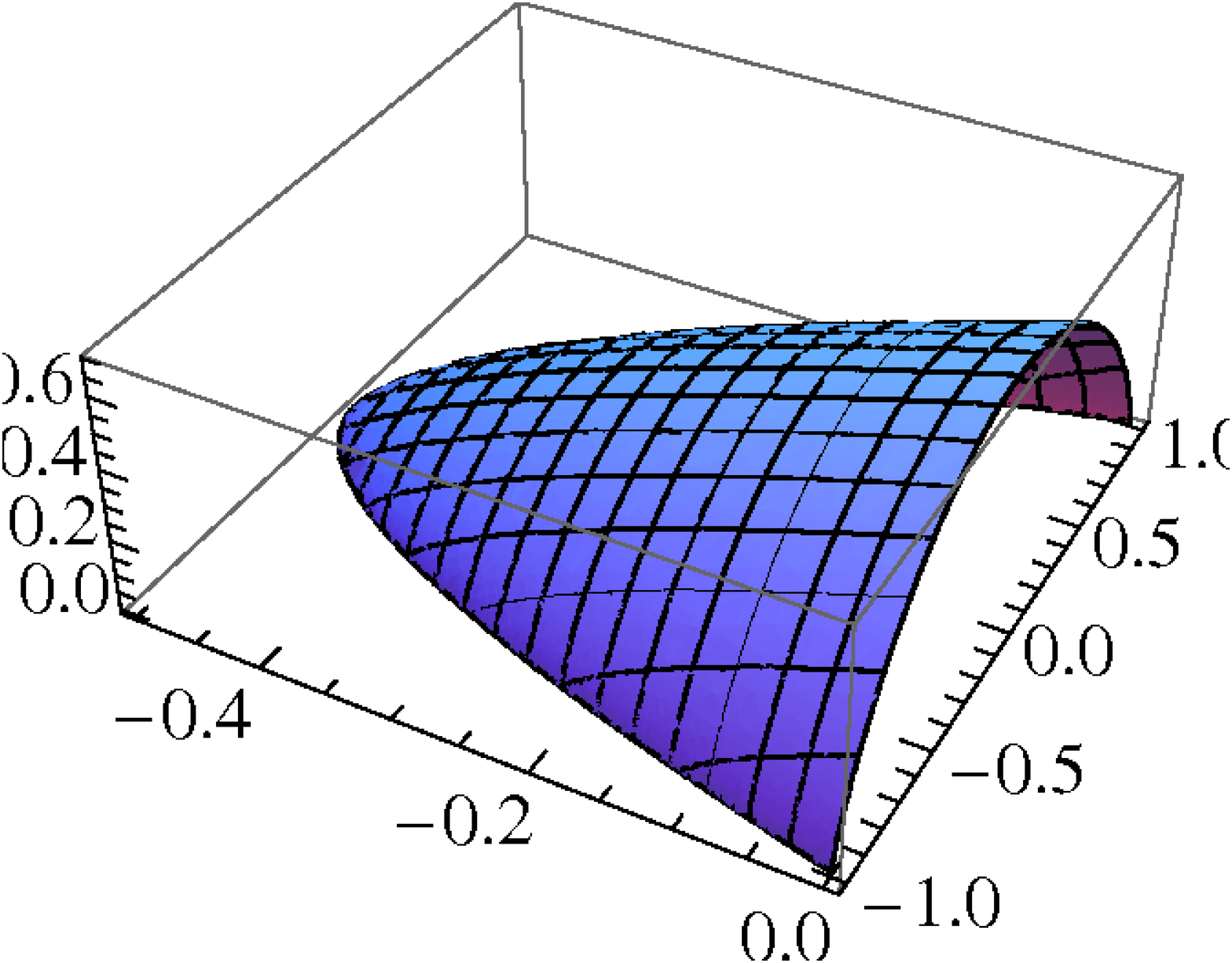}}
\caption{\label{fig:entropy}
(Color online) Domains (left) and graphs (right) of the microcanonical entropy $s(e,m)$ of the anisotropic quantum Heisenberg model for some combinations of the couplings $\lambda_\perp,\lambda_3$. From top to bottom: $(\lambda_\perp,\lambda_3)=(1/4,1)$, $(9/10,1)$, $(1,1)$, $(1,9/10)$, $(1,1/2)$, $(1,1/5)$, $(1,0)$. For the domains, the abscissa is the energy $e$ and the ordinate is the magnetization $m$, and the entropy is defined on the shaded area.
}
\end{figure}

The final result for the microcanonical entropy of the Curie-Weiss anisotropic quantum Heisenberg model in the thermodynamic limit is
\begin{equation}\label{eq:sem}
\begin{split}
s(e,m)=&\ln2-\frac{1}{2}[1-f(e,m)]\ln[1-f(e,m)]\\
&-\frac{1}{2}[1+f(e,m)]\ln[1+f(e,m)]
\end{split}
\end{equation}
with
\begin{equation}\label{eq:fem}
f(e,m)=\sqrt{m^2\left(1-\frac{\lambda_3}{\lambda_\perp}\right)-\frac{2e}{\lambda_\perp}},
\end{equation}
and $\lambda_\perp=\max\{\lambda_1,\lambda_2\}$ \footnote{$\lambda_\perp=0$ has to be considered separately, but this is just the well-known case of the Curie-Weiss Ising model.}, where $s(e,m)$ is defined on the subset of $\RR^2$ for which
\begin{equation}\label{eq:domain}
0<m^2(\lambda_\perp-\lambda_3)-2e<\lambda_\perp\quad\text{and}\quad 2e<-m^2\lambda_3.
\end{equation}
The result is remarkably simple, in the sense that an explicit expression for $s(e,m)$ can be given. This is in contrast to the canonical ensemble, where $g(\beta,h)$ is given implicitly as the solution of a maximization \cite{PearceThompson75}. Plots of the domains and graphs of $s(e,m)$ are shown in Fig.\ \ref{fig:entropy} for a number of coupling strengths $\lambda_\perp$, $\lambda_3$.
%------------------------------------------------
%------------------------------------------------

{\em Nonequivalence of ensembles.---} On a thermodynamic level, equivalence or nonequivalence of the microcanonical and the canonical ensembles is related to the concavity or nonconcavity of the microcanonical entropy \cite{TouElTur04}. By inspection of rows three to seven in Fig.\ \ref{fig:entropy} [or by simple analysis of the results in \eqref{eq:sem}--\eqref{eq:domain}], the entropy $s$ for $\lambda_\perp>\lambda_3$ is seen to be a concave function on a domain which is a convex set. For $\lambda_\perp<\lambda_3$, the domain is not a convex set and therefore the entropy is neither convex nor concave. In the latter case, microcanonical and canonical ensembles are not equivalent, in the sense that it is impossible to obtain the microcanonical entropy $s(e,m)$ from the canonical Gibbs free energy $g(\beta,h)$, although the converse is always possible by means of a Legendre-Fenchel transform.

The physical interpretation of ensemble equivalence is that every thermodynamic equilibrium state of the system that can be probed by fixing certain values for $e$ and $m$ can also be probed by fixing the corresponding values of the inverse temperature $\beta(e,m)$ and the magnetic field $h(e,m)$. In the situation $\lambda_\perp<\lambda_3$ where nonequivalence holds, this is not the case: only equilibrium states corresponding to values of $(e,m)$ for which $s$ coincides with its concave envelope can be probed by fixing $(\beta,h)$; macrostates corresponding to other values of $(e,m)$, however, are not accessible as thermodynamic equilibrium states when controlling temperature and field in the canonical ensemble. In this sense, microcanonical thermodynamics can be considered not only as different from its canonical counterpart, but also as richer, allowing to probe equilibrium states of matter which are otherwise inaccessible. The realization of a long-range quantum spin system by means of a cold dipolar gas in an optical lattice offers the unique and exciting possibility to study such states in a fully controlled laboratory setting \footnote{There is a further long-range peculiarity, going under the name of partial equivalence \cite{CaKa07}, which can be observed in the anisotropic quantum Heisenberg model for any values of the coupling constants $\lambda_1$, $\lambda_2$, and $\lambda_3$. Partial equivalence here refers to the situation that a macrostate, associated with a certain pair of values $(e,m)$ in the microcanonical ensemble, corresponds to more than just one pair of values $(\beta,h)$ canonically.}.

%------------------------------------------------

{\em Thermodynamic equivalence of models.---} Let us leave aside for a moment the question of experimental realization and discuss a different kind of equivalence specific to the anisotropic quantum Heisenberg model. It had been observed already in the 1970s that the isotropic Heisenberg model and the Ising model are thermodynamically equivalent in the sense that their canonical free energies coincide \cite{Niemeyer70}. One can verify by Legendre-Fenchel transforming the entropy in \eqref{eq:sem} that the same is in fact true for all coupling strengths satisfying $\lambda_\perp\leqslant\lambda_3$. Geometrically, this thermodynamic equivalence corresponds to the fact that the entropies $s(e,m)$ for those couplings share the same concave hull (which is equal to the entropy of the isotropic Heisenberg model plotted in row three of Fig.\ \ref{fig:entropy}). Identical concave hulls of entropies imply, however, identical canonical free energies, and hence thermodynamic equivalence of the two models follows in the canonical ensemble. Remarkably, however, thermodynamic equivalence does not hold in the microcanonical ensemble, as is obvious from the different shapes of entropies in rows one to three of Fig.\ \ref{fig:entropy}.

%------------------------------------------------

%{\em Discussion.---} The microcanonical entropies \eqref{eq:semdef} and \eqref{eq:sNemdef} discussed in this Letter describe the physical situation of fixed energy $e$ and fixed magnetization $m$, but it is not immediately clear how to realize a conserved magnetization experimentally. A way out consists in a bosonic lattice gas interpretation of spin systems, as pioneered by Matsubara and Matsuda \cite{MatMat56}, in which the magnetization of a spin system corresponds to the particle density $\varrho$ on a lattice via the relation $\varrho=m+1/2$, and this density is usually conserved to a very good degree in an optical lattice experiment. A similar, but classical, lattice gas interpretation was proposed earlier for an optical lattice-realization of an Ising system in \cite{KaPlei09}.

{\em Discussion.---} The microcanonical entropies \eqref{eq:semdef} and \eqref{eq:sNemdef} discussed in this Letter describe the physical situation of fixed energy $e$ and magnetization $m$. In a cold atom experiment, energy is conserved to a very good degree due to the absence of a heat bath. For apolar gases where $s$-wave scattering is dominant, the total magnetization is also fixed, and the resulting short-range interacting microcanonical spin systems have been discussed in \cite{KaPlei09}. For dipolar gases where long-range interactions are present and nonequivalent ensembles can occur, the magnetization is not conserved in general (unless an experimentalist comes up with an ingenious trick the author is not aware of). More easily, nonequivalence of ensembles could be observed in long-range quantum spin systems undergoing a temperature-driven first-order transition. In this case, nonequivalence is signalled by a nonconcave microcanonical entropy $s(e)$, corresponding to conservation of energy $e$, but fluctuating magnetization.

Although the anisotropic quantum Heisenberg model discussed in this Letter is among the systems which can be engineered with cold polar molecules in optical lattices \cite{Micheli_etal06}, this model is chosen here not for its particular features, but to illustrate general, and possibly even generic, properties of long-range interacting quantum spin systems: nonconcave entropies, nonequivalence of statistical ensembles, or other phenomena like negative microcanonical response functions must be expected to show up under the experimental conditions realized in experiments with cold dipolar gases in optical lattices in general. Finally, note that dipolar atoms or molecules are not the only possible realization of long-range quantum spin systems in optical lattices: Following a suggestion by O'Dell {\em et al.} \cite{ODell_etal00}, long-range interactions decaying with the interparticle distance $r$ like $r^{-1}$ can be engineered by shining appropriately tuned laser light onto atoms, even in the absence of a permanent dipole moment.

{\em Summary.---} A calculation of the microcanonical entropy of the anisotropic Curie-Weiss quantum Heisenberg model was reported. The results illustrate peculiarities of long-range quantum spin systems, like nonconcave entropies and nonequivalence of statistical ensembles. The microcanonical setting models the conditions relevant for experiments with dipolar gases in optical lattices. The results point out the importance of nonstandard thermodynamics beyond the canonical ensemble for such experiments on the one hand, and on the other hand suggest the use of optical lattice experiments for the study of fundamental issues of thermostatistics.

\bibliography{MFQuantumSpin.bib}

\end{document}